**REVIEW**                                                                                      **Open Access**

# Update on the genetic and epigenetic etiology of gestational diabetes mellitus: a review

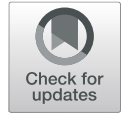

Tajudeen O. Yahaya[1*], Titilola Salisu[2], Yusuf B. Abdulrahman[3] and Abdulrazak K. Umar[3]

## Abstract

**Background:** Many studies have been conducted on the genetic and epigenetic etiology of gestational diabetes mellitus (GDM) in the last two decades because of the disease's increasing prevalence and role in global diabetes mellitus (DM) explosion. An update on the genetic and epigenetic etiology of GDM then becomes imperative to better understand and stem the rising incidence of the disease. This review, therefore, articulated GDM candidate genes and their pathophysiology for the awareness of stakeholders.

**Main body (genetic and epigenetic etiology, GDM):** The search discovered 83 GDM candidate genes, of which *TCF7L2*, *MTNR1B*, *CDKAL1*, *IRS1*, and *KCNQ1* are the most prevalent. Certain polymorphisms of these genes can modulate beta-cell dysfunction, adiposity, obesity, and insulin resistance through several mechanisms. Environmental triggers such as diets, pollutants, and microbes may also cause epigenetic changes in these genes, resulting in a loss of insulin-boosting and glucose metabolism functions. Early detection and adequate management may resolve the condition after delivery; otherwise, it will progress to maternal type 2 diabetes mellitus (T2DM) and fetal configuration to future obesity and DM. This shows that GDM is a strong risk factor for T2DM and, in rare cases, type 1 diabetes mellitus (T1DM) and maturity-onset diabetes of the young (MODY). This further shows that GDM significantly contributes to the rising incidence and burden of DM worldwide and its prevention may reverse the trend.

**Conclusion:** Mutations and epigenetic changes in certain genes are strong risk factors for GDM. For affected individuals with such etiologies, medical practitioners should formulate drugs and treatment procedures that target these genes and their pathophysiology.

## Background

Pregnant women develop insulin resistance at certain stages owing to increased placenta hormones, but most women overcome this condition by up-regulating insulin production through beta cell expansion [1]. Gestational diabetes mellitus (GDM) begins when a pregnant female does not make the extra insulin needed to normalize blood glucose during the second or third trimester of pregnancy [1]. Sometimes the glucose intolerance may be present before pregnancy, but not diagnosed [2, 3]. Uncontrolled GDM can cause high blood pressure, type

2 diabetes mellitus (T2DM), and increased risks of vascular diseases in pregnant women [4, 5]. Intrauterine exposure to high blood glucose may program the offspring to develop diabetes or obesity later in life [6]. It may also cause macrosomia, birth defects, preterm birth, and developmental delay [7–9].

GDM is the most common metabolic condition during pregnancy [10] with a global incidence rate between 1 and 28 % [11]. In 2017, GDM affected about 204 million women worldwide with a projection to increase to 308 million by 2045, mostly in developing countries [12]. Though most times the glucose intolerance normalizes soon after delivery, women with GDM have a high risk of developing T2DM later in life [13]. Compared to women with normal glucose tolerance, women with

* Correspondence: yahayatajudeen@gmail.com; yahaya.tajudeen@fubk.edu.ng
[1]Federal University Birnin Kebbi, PMB 1157 Birnin Kebbi, Nigeria
Full list of author information is available at the end of the article

GDM are at least seven times more predisposed to T2DM [14]. Moreover, almost half of pregnant females with GDM will develop diabetes in a decade [14]. Offspring of women with GDM are also 8 times more prone to diabetes or pre-diabetes [15]. These show that GDM contributes immensely to the alarming incidence of diabetes worldwide [16]. Diabetes affected about 451 million people in 2017, of which 5 million died and USD 850 billion was spent on healthcare expenditure [17]. Considering the impact of GDM, the reduction of its prevalence and effective management of the affected will go a long way in stemming the incidence and burden of diabetes. However, to achieve a reduced prevalence of GDM, a proper understanding of its etiology is necessary. Fortunately, improved biological techniques in the last two decades have led to more understanding of the genetic and epigenetic etiology of the disease, thus an update becomes necessary. This review therefore articulated current findings on the genetic and epigenetic etiology of GDM.

## Methods
### Databases searched
An in-depth search of PubMed, Scopus, SpringerLink, Google Scholar, and ResearchGate databases was performed for relevant research articles on GDM.

### Search terms
Some search terms used to retrieve articles are gestational diabetes mellitus, hyperglycemia, insulin resistance, obesity, glucose metabolism, beta-cell dysfunction, and gestational diabetes genes. Other search terms used include glucose insensitivity, epigenetics of diabetes mellitus, gestational diabetes testing and cost-effectiveness, and the prevalence of gestational diabetes.

### Article inclusion criteria
Inclusion criteria include the following:
   Research published in the English language.
   Research that focused on GDM.
   Studies that focused on the genetic and epigenetic etiology of GDM.
   Articles that centered on GDM testing and cost-effectiveness.
   Studies published between 2000 till date.

### Article exclusion criteria
Exclusion criteria include the following:
   Studies that are not available in English language.
   Studies with only abstract available.
   Research that described GDM, but with no clear genetic and epigenetic mechanisms
   Studies published before the year 2000.

## Results
### Genetic etiology in GDM
The search found that mutations in some genes, or their variants, may interact with one another and environmental triggers to cause GDM (Fig. 1). The genetic etiology of GDM overlaps with T2DM, as most of the GDM candidate genes also predispose humans to T2DM. This explains the common pathophysiology of GDM and T2DM as both express beta-cell dysfunction and abnormal glucose metabolism [18]. Contrary to some reports, GDM also shares common pathophysiology with type 1 diabetes mellitus (T1DM) and maturity-onset diabetes of the young (MODY), but less than 10 % of GDM patients show these associations [18].

   Though many genes reportedly showed an association with obesity, insulin resistance, and beta-cell dysfunction, the present study discovered only 83 genes with a clear GDM pathophysiology and are presented in Table 1.

### Most frequent GDM candidate genes
The list of GDM candidate genes is inexhaustible as more genes are continually discovered; however, certain genes are most often linked with the disease. Table 2 shows the most frequent GDM candidate genes and their variants in various ethnic groups.

   From Table 2, we used a pie chart (Fig. 2) to express the percentage occurrence of each gene based on ethnicity. *TCF7L2* gene was the most frequent having present in 17 % of the ethnics, followed by *MTNR1B* with 15 %, *CDKAL1* 10 %, *KCNQ1* 10 %, and *IRS1* 10 %. Some other genes are not widespread, but are often found in certain ethnics or regions. These genes include *ZRANB3*, found in Africa [96]; *ABCC8* found among Finnish [119]; *Chemerin*, found among Iranians [120]; and *INS*, found among the Greeks [121]. These genes can be used to develop a genetic testing guideline to predict the likelihood of GDM or determine its genetic and epigenetic etiology. This is important because there is no genetic testing procedure yet for GDM, partly because the condition is multifactorial in which several genes interact with environmental triggers to cause the disease. Thus, a mutation in a single gene may not explain full susceptibility to GDM and testing for all the candidate genes will be expensive and cumbersome. The low prevalence of GDM in the past also contributes to the lack of interest in developing a genetic testing guideline for the disease.

### Epigenetic etiology in GDM
Epigenetics refers to the study of heritable changes in biological processes caused by modification of chemical tags on DNA such as methyl and ethyl groups [122]. These modifications are mediated by some mechanisms, including DNA methylation, histone modification, and microRNA expression [122]. Epigenetic mechanisms play important roles in several cellular activities, but certain environmental triggers can reprogram the epigenome, resulting in disease pathologies [123]. In particular, epigenetic mechanisms regulate several genes that maintain beta-cell morphology, proliferation, and functions, thus implies that epigenetic modification may disrupt insulin secretion and sensitivity, causing metabolic diseases, including GDM [124, 125]. Epigenetic changes are expressed in both somatic and gamete cells, and thus can be transmitted from generation to generation [126].

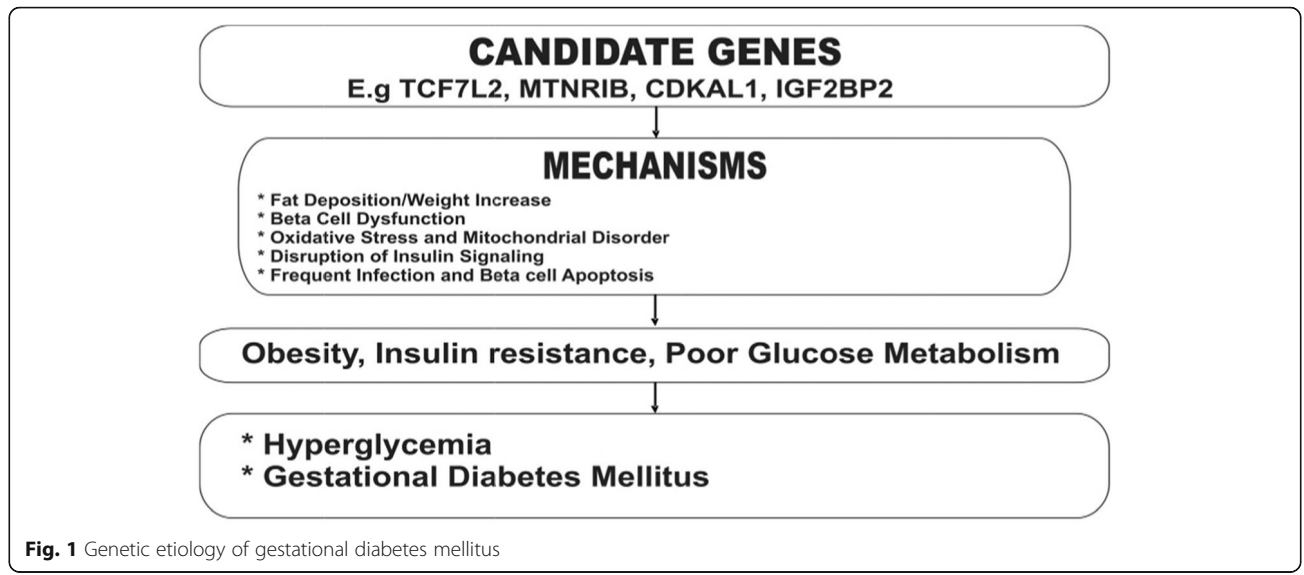

**Fig. 1** Genetic etiology of gestational diabetes mellitus

Studies have reported many instances of epigenetic modifications involving insulin synthesis and glucose metabolism in GDM. For instance, histone under-acetylation and overmethylation in the promoter region of the *PDX1* gene reduce the insulin-boosting function of the gene [127]. Also, in a study that examined the methylation of the *IL-10* gene among pregnant women, hypo-methylation of maternal blood cells and elevated plasma *IL-10* levels were noticed in women with GDM [128]. In another study that compared the *miRNA* profiles of some diabetic pregnant rats with nondiabetic, repression of *miR-338* and overexpression of *miR-451* were associated with reduced β-cell mass in the diabetic [129]. In vitro upregulation of *miR-451* and repression of miR-338 in the same study increased β-cell mass, leading to improved glucose metabolism [129]. In a study that investigated the *miRNA* expressions of maternal and fetal blood cells of some pregnant women, 29 *miRNAs* were upregulated in individuals with GDM [130]. Of these *miRNAs*, *miRNA-340* was confirmed to downregulate the expression of the *PAIP1* gene [130]. In vitro normalization of the *miRNA-340* expression of the diabetic mothers increased insulin production [130].

## Environmental triggers of genetic and epigenetic etiology of GDM

Both the genetic and epigenetic etiology of GDM are mediated by certain environmental triggers that change gene functions. The genetic triggers mutate the genes, while the epigenetic triggers affect the chemical tags on the DNA without affecting the nucleotide sequence (Fig. 3). Among the environmental triggers are pollution and microbial exposures, whose GDM modulatory roles have been established by several studies. In a study that monitored the effects of air pollution among pregnant Southern Californians, prepregnancy exposures to nitrogen dioxide ($NO_2$), particulate matter (PM2.5 and PM10), and dioxin were related to GDM [131]. Nitrogen dioxide and particulate matter can cause oxidative stress, overexpression of proinflammatory cytokines, and endothelial dysfunction, resulting in increased insulin resistance [132]. Dioxin compounds can interact with peroxisome proliferator-activated receptor-γ (*PPARG*), disrupting insulin signaling pathways and resulting in insulin resistance and abnormal glucose metabolism [133]. Exposure to pathogenic microbial organisms may disrupt the gut microbiota and compromise the immune system, leading to metabolic disorders and GDM. Vu *et al.* [134] demonstrated in rabbits that a toxin produced by *Staphylococcus aureus* may interact with fat cells and the immune system, resulting in inflammation, insulin resistance and glucose intolerance [134]. In a study of the microbiota of some pregnant women, individuals with GDM showed gut microbiota imbalance containing majorly the phylum Actinobacteria and the genus Collinsella, Rothia and Desulfovibrio [135]. A balanced gut microbiota is necessary for optimum metabolism and the immune system. Aside from microbial infection, other environmental factors that may disrupt

**Table 1** GDM predisposing genes showing locations and phathophysiology

| Number | Gene | Full name | Locus | Pathophysiology |
|---|---|---|---|---|
| 1 | TCF7L2 | Transcription factor 7-like 2 | 10q2 | Increases apoptosis, impairing insulin secretion [19]. |
| 2 | KCNQ1 | Potassium voltage-gated channel sub-family Q member 1 | 11p15.5-15.4 | It disrupts the influx of calcium into the channel, resulting in decreased insulin secretion [20]. |
| 3 | CENTD2/ ARAP1 | Centaurin-delta-2/ ArfGAP with rhoGAP domain, ankyrin repeat and PH domain 1 | 11q13.4 | Causes disruption of glucose-induced insulin secretion [21]. |
| 4 | MTNR1B | Melatonin receptor 1B | 11q14.3 | Decreases insulin secretion, elevating fasting glucose levels [22]. |
| 5 | IGF1 | Insulin-like growth factor 1 | 12q23.2 | Induces high body mass (HBM), leading to metabolic disturbances, especially insulin resistance and hyperinsulinemia [23]. |
| 6 | IGF2 | Insulin-like growth factor 2 | 11p15.5 | Overexpression of IGF2 leads to β-cell dedifferentiation and endoplasmic reticulum stress, causing islet dysfunction [24]. |
| 7 | IGFBP-1 | Insulin like growth factor binding protein 1 | 7p12.3 | Decreased blood levels of IGFBP-1 cause overexpression of IGF-I, resulting in inflammation [25]. |
| 8 | IGFBP-2 | Insulin like growth factor binding protein 3 | 2q35 | Reduced expression of IGFBP-2 inhibits adipogenesis, leading to obesity and insulin resistance [26]. |
| 9 | IGF2BP2 | Insulin like growth factor 2 mRNA binding protein 2 | 3q27.2 | Impairs β-cell function and modulates obesity, altering insulin sensitivity [27]. |
| 10 | IGFBP-3 | Insulin like growth factor binding protein 3 | 7p12.3 | Overexpression of IGFBP-3 predisposes to HBM body, disrupting glucose metabolism [28]. |
| 11 | IGFBP-4 | Insulin like growth factor binding protein 4 | 17q21.2 | Reduced levels cause HBM and insulin resistance [28]. |
| 12 | IGFBP5 | Insulin like growth factor binding protein 5 | 2q35 | Disrupts IGF-1 signaling pathway, leading to insulin insensitivity [29]. |
| 13 | PPARG | Peroxisome proliferator-activated receptor gamma | 3p25.2 | Stimulates abnormal fat deposition in tissues, causing obesity and insulin resistance [27]. |
| 14 | KCNJ11 | Potassium voltage-gated channel sub-family J member 11 | 11p15.1 | Reduces the sensitivity of pancreatic beta-cell KATP channel subunit (Kir6.2), resulting in decreased insulin release [28]. |
| 15 | INSR | Insulin receptor | 19p13.2 | Predisposes to obesity, leading to insulin resistance [29]. |
| 16 | ADRB2 | Adrenoceptor beta 2 | 5q32 | Increases the secretion of vascular endothelial growth Factor-A (VEGF-A) in the β-cells, resulting in hyper-vascularized islets and disrupting insulin secretion and glucose metabolism [30]. |
| 17 | ADRB3 | Adrenoceptor beta 3 | 8p11.23 | Increases body weight, predisposing to obesity and insulin resistance [31]. |
| 18 | GNB3 | G protein subunit beta 3 | 12p13.31 | Causes high-fat deposition and obesity [32]. |
| 19 | ABCC8 | ATP binding cassette subfamily c member 8 | 11p15.1 | Loss of function of the gene disrupts the KATP channel function, increasing the body weight and causing hyperinsulinism [33]. |
| 20 | CAPN10 | Malpain 10 | 2q37.3 | Increases body mass, initiating insulin resistance [34]. |
| 21 | MBL2 | Mannose-binding lectin | 10q21.1 | Causes frequent infections and chronic inflammatory diseases, leading to high-fat deposition and insulin resistance [35]. |
| 22 | GLUT4/ SLC2A4 | Glucose transporter type 4/Solute carrier family 2 member 4 | 17p13.1 | Impairs insulin signaling pathway [36]. |
| 23 | RBP4 | Retinol binding protein-4 | 10q23.33 | Increases gluconeogenesis and impairs insulin signaling in muscles [37]. |
| 24 | PCK1 | Phosphoenolpyruvate carboxykinase 1 | 20q13.31 | Induces high levels of fasting insulin, causing abnormal glucose metabolism [38]. |
| 25 | PIK3R1/ PI3K | Phosphoinositide-3-kinase regulatory subunit 1 | 5q13.1 | Disrupts insulin signaling pathway in skeletal muscle and inhibit liver gluconeogenesis [38]. |
| 26 | STRA6 | Signaling receptor and transporter of retinol STRA6 | 15q24.1 | Promotes fat deposition, predisposing to obesity and insulin resistance [39]. |
| 27 | VDR | Vitamin D receptor | 12q13 | Predisposes to obesity, causing metabolic disorder, especially insulin resistance [40, 41]. |
| 28 | CDKAL1 | Cyclin-dependent Kinase 5 Regulatory subunit-associated protein 1-like 1 | 6p22.3 | Inhibits the conversion of proinsulin to insulin through protein translation, leading to insulin resistance [42]. |
| 29 | GCK | Glucokinase | 7p13 | Increases body fat mass, resulting in insulin resistance [43]. |

**Table 1** GDM predisposing genes showing locations and phathophysiology (Continued)

| Number | Gene | Full name | Locus | Pathophysiology |
|---|---|---|---|---|
| 30 | CDKN2A/2B | Cyclin-dependent kinase inhibitor 2a | 9p21.3 | Affects proinsulin conversion to insulin and reduces insulin sensitivity [44]. |
| 31 | SRR | Serine racemase | 17p13.3 | Disrupts the secretion of insulin and/or glucagon [45]. |
| 32 | HHEX/IDE | Hematopoietically expressed homeobox | 10q23.33 | Causes pancreatic and liver developmental error [46]. |
| 33 | SLC30A8 | Solute carrier family 30 member 8 | 8q24.11 | Modulates loss of zinc in the beta cells, destabilizing insulin molecules[47]. |
| 34 | LEP | Leptin | 7q31.3 | Promotes inflammation, causing energy imbalance and obesity [48]. |
| 35 | LEPR | leptin receptor | 1p31 | Induces high-fat mass and insulin resistance [49]. |
| 36 | HNF1B/TCF2 | Hepatocyte nuclear factor 1B | 17q12 | Causes β-cell dysfunction [50, 51]. |
| 37 | TNF-α/TNF | Tumor necrosis factor-α | 6p21.33 | Causes inflammatory and oxidative stress [52]). |
| 38 | HNF4A/TCF1 | Hepatocyte nuclear factor 4 alpha | 20q12 | Induces β-cell dysfunction [50, 51]. |
| 39 | WFS1 | Wolfram syndrome 1 | 4p16.1 | Initiates endoplasmic reticulum stress and mitochondrial disorder, leading to β-cell dysfunction [53]. |
| 40 | IRS1 | insulin receptor substrate 1 | 2q36.3 | Induces an inflammatory response and causing low insulin sensitivity [54]. |
| 41 | HTR2B/ 5-HT-1A | 5-hydroxytryptamine receptor 1a | 5q12.3 | Reduces beta-cell proliferation and increases body weight [55]. |
| 42 | TPH1 | Tryptophan hydroxylase 1 | 11p15.1 | Causes low levels of serotonin, increasing weight gain and causing insulin intolerance [56]. |
| 43 | 5-HT1A/ HTR3A | 5-hydroxytryptamine receptor 1a | 5q12.3 | Causes low serotonin levels, resulting in insulin resistance [57]. |
| 44 | HNF1A | Hepatocyte nuclear factor-1 alpha | 12q24.31 | Causes adiposity, leading to pre-pregnancy obesity and insulin resistance [57]. |
| 45 | GCKR | Glucokinase regulator | 2p23.3 | Overexpression of GCKR causes hyperactivity of GCK, reducing glucose and increasing fat accumulation [58]. Loss of the function reduces GCK expression, impairing glucose clearance [59]. |
| 46 | MIF | Macrophage migration inhibitory factor | 22q11.23 | Overexpression of the MIF gene in placental tissue causes insulin resistance [60]. |
| 47 | ADRA2A | Alpha-2-adrenergic receptors | 10q25.2 | Increases body fat mass, leading to loss of glucose regulation [61]. |
| 48 | SLC6A4 | Solute carrier family 6 member 4 | 17q11.2 | Impairs serotonin metabolism, increasing body weight and causing insulin resistance [62]. |
| 49 | FTO | Fat mass and obesity-associated gene/ Alpha-ketoglutarate dependent dioxygenase | 16q12.2 | Causes adiposity, leading to pre-pregnancy obesity and insulin resistance [63]. |
| 50 | TLE1 | Transducin-like enhancer of split-1 | 9p21.32 | Elevates fasting glucose level and reduces insulin secretion [64]. |
| 51 | ADCY5 | Adenylate cyclase 5 | 3q21.1 | Alters ADCY5 expression in pancreatic beta cells, impairing glucose signaling [65]. |
| 52 | IL-1β | Interleukin-1 beta | 2q14.1 | Impairs pancreatic β-cells, decreasing insulin secretion [66]. |
| 53 | IL-6 | Interleukin-6 | 7p15.3 | Overexpression destroys pancreatic β-cells, resulting in apoptosis and low insulin synthesis [67]. |
| 54 | IL-10 | Interleukin-10 | 1q32.1 | Overexpression compromises immune response, disrupting insulin metabolism [68]. |
| 55 | PAX8 | Paired box 8 | 2q14.1 | Reduces islet viability and beta cell survival [69]. |
| 56 | ADIPOQ (diponectin gene) | Adiponectin, C1Q and collagen domain containing | 3q27.3 | Causes low adiponectin, leading to obesity and insulin resistance [70]. |
| 57 | RARRES2 (Chemerin gene) | retinoic acid receptor responder 2 | 7q36.1 | Initiates inflammation and energy imbalance, leading to obesity and insulin resistance [71]. |
| 58 | SERPINA12 (Vaspin gene) | Serpin family a member 12 | 14q32.13 | Causes inflammation, loss of energy balance, and obesity [72]. |
| 59 | RETN | Resistin | 19p13.2 | Causes a loss of energy balance, obesity, and insulin resistance [73]. |

**Table 1** GDM predisposing genes showing locations and phathophysiology *(Continued)*

| Number | Gene | Full name | Locus | Pathophysiology |
|---|---|---|---|---|
| 60 | APLN | Apelin | Xq26.1 | Causes a loss of energy balance, obesity, and insulin resistance [74]. |
| 61 | NUCB2 (nesfatin 1 gene) | Nucleobindin 2 | 11p15.1 | Causes a loss of energy balance, obesity, and insulin resistance [75]. |
| 62 | ITLN1 | Intelectin-1/Omentin-1 | 1q23.3 | Loss of function induces insulin resistance [76]. |
| 63 | NAMPT/PBEF1 (Visfatin gene) | Nicotinamide phosphoribosyltransferase | 7q22.3 | Causes obesity and insulin resistance [76]. |
| 64 | HMG20A/ iBRAF | High mobility group protein 20a | 15q24.3 | Depletion represses expression of insulin-producing genes such as Neu-roD, Mafa and GCK, and enhances beta-cell de-differentiating gene such as PAX4 and REST [77]. |
| 65 | RREB1 | Ras responsive element binding protein 1 | 6p24.3 | Causes fat deposition and beta cell dysfunction [78]. |
| 66 | GLIS3 | GLIS family zinc finger 3 | 9p24.2 | Causes fat deposition and beta cell dysfunction [78]. |
| 67 | GPSM1 | G protein signaling modulator 1 | 9q34.3 | Causes fat deposition and beta cell dysfunction [78]. |
| 68 | mtDNA | Mitochondrial DNA | All cells | Induces oxidative stress and mitochondrial disorder, causing insulin resistance [79]. |
| 69 | PRLR | Prolactin receptor | 5p13.2 | Modulates loss of PRLR signaling in β-cells. reducing β-cell proliferation and expansion during pregnancy [80]. |
| 70 | MAFB | MAF bZIP transcription factor B | 20q12 | Causes inadequate β-cell expansion [80]. |
| 71 | SERT | Serotonin transporter | 17q11.1-12 | Stimulates abnormal fat accumulation in both white and brown adipose tissues, causing glucose intolerance and insulin resistance [81]. |
| 72 | PAI-1 | Plasminogen activator inhibitor 1 | 7q22 | Predisposes to adiposity, increasing body weight and affecting pancreatic beta-cell function [82]. |
| 73 | TSPAN8 | Tetraspanin-8 | 12q21.1 | Impairs gestational glucose tolerance [83]. |
| 74 | G6PC2 | Glucose-6-phosphatase catalytic subunit 2 | 2q31.1 | Elevates fasting glucose level and reduces insulin secretion [64]. |
| 75 | PTPRD | Protein tyrosine phosphatase receptor type d | 9p24.1-p23 | Disrupts insulin signaling pathway, leading to altered insulin sensitivity and glucose homeostasis [84]. |
| 76 | CRP | C-reactive protein | 1q23.2 | Overexpression causes obesity, resulting in systemic inflammation and insulin resistance [85]. |
| 77 | GK | Glycerol kinase | Xp21.2 | Deficiency causes abnormal insulin metabolism [86]. |
| 78 | PAX4 | Paired box gene 4 | 7q32.1 | Impairs fetal islet cell differentiation, altering insulin sensitivity later in life [87]. |
| 79 | HDAC4 | Histone deacetylase 4 | 2q37.3 | Causes β-cell loss, leading to decreased insulin secretion. Also represses beta-cell transcriptional factors [88]. |
| 80 | FETUA/ AHSG | Fetuin-a | 3q27.3 | Increases body mass, insulin secretion and C-peptide levels, but lower insulin sensitivity [89]. |
| 81 | FETUB | Fetuin-b | 3q27.3 | Increases hepatic steatosis, impairing insulin secretion and glucose metabolism [90]. |
| 82 | FGF21 | Fibroblast growth factor 21 | 10q26.13 | Cause an abnormal glucose metabolism independent of insulin resistance [91]. |
| 83 | SNORA8 | | | An emerging candidate gene [92] |

gut microbiota include certain diseases, oral microbiome, diets, and antibiotic use, among others [136].

Lifestyles such as short sleep, poor nutritional choices, advanced age, and physical inactivity are some environmental triggers that may predispose humans to GDM. Short hour sleep at night can increase body fat accumulation, reduce glucose metabolism, and predispose to T2DM and GDM. In a study that determines the frequency of GDM among sleep-deprived 668 Singaporeans, 131, representing 19 %, were diagnosed with GDM of which 27.3 % sleep less than 6 hours a night, while 16.8 % sleep between 7-8 hours [137]. Poor nutrition, such as energy-dense western diets may cause overweight and obesity, disrupting insulin signaling pathways and insulin sensitivity. Saturated fats can disrupt insulin signaling, induce inflammation and endothelial dysfunction, resulting in GDM. In a study that evaluated the effect of dietary patterns among pregnant Chinese women, diets containing high protein and low starch were associated with a reduced risk of GDM [138]. Energy-dense diets are deficient in betaine, which is a methyl donor for methylating important biological processes and as well a substrate of methionine metabolism [139]. Diets low in betaine may induce abnormal methylation of some genes involved in insulin synthesis and glucose metabolism. Advanced age may also predispose pregnant women to GDM because mitochondrial functions decline with age, leading to reduced metabolic activities and an increased body mass index [140]. Aging changes the epigenetic pattern, affecting the expression of some genes involved in glucose metabolism, particularly the *COX7A1* gene in the respiratory chain [141]. In a study of 1688 women in northwest London who developed GDM, advanced maternal age was linked to GDM [142]. Studies that demonstrate the role of physical inactivity in the pathogenesis of GDM are scarce. However, a systematic and meta-analysis by Ming *et al.* [143] shows that physical activity during pregnancy can decrease the occurrence of GDM, suggesting that lack of exercise is a risk factor. Some mechanisms through which inactivity mediates diseases include mitochondrial dysfunction, changes in the composition of muscles, and insulin resistance, among others [144]. Physical inactivity influences the epigenome negatively, affecting several generations [145].

**Table 2** Most prevalent gestational diabetes mellitus genes across countries, ethnicity, and race in the world

| Gene | Countries/ethnicity/race | Variants | Level of significance | References |
|---|---|---|---|---|
| TCF7L2 | Mexicans | rs7901695 | $P = 2.16 \times 10^{-6}$ | [93] |
| | | rs7901696 | $P < 0.05$ | [93] |
| | | rs7901697 | $P < 0.05$ | [93] |
| | | rs7901698 | $P < 0.05$ | [93] |
| | Asians | rs7903146 | $P = 0.001$ | [94] |
| | Scandinavians | rs7903146 | $P < 0.05$ | [95] |
| | Africans | rs7903146 | $P = 7.288 \times 10^{-13}$ | [96] |
| | Hispanic/Latinos | rs7903146 | $P < 0.05$ | [97] |
| | Caucasians/Danish | rs7903146 | $P = 0.00017$ | [98] |
| | Caucasian | rs4506565 | $P < 0.001$ | [99] |
| KCNQ1 | Mexicans | rs2237892 | $P = 1.98 \times 10^{-5}$ | [93] |
| | | rs163184 | $P < 0.05$ | [93] |
| | | rs2237897 | $P < 0.05$ | [93] |
| | Koreans | rs2074196 | $P = 0.039$ | [100] |
| | | rs2237892 | $P < 0.05$ | [100] |
| | East Asians | rs2074196 | $P = 0.039$ | [100] |
| | | rs2237892 | $P < 0.05$ | [100] |
| | Pakistan | rs2237895 | $P < 0.05$ | [101] |
| MTNR1B | Mexicans | rs1387153 | $P = 0.05358$ | [93] |
| | Asians | rs10830963 | $P < 0.001$ | [94] |
| | Caucasians | rs10830963 | $P < 0.001$ | [94] |
| | Koreans | rs10830962 | $P = 2.49 \times 10^{-13}$ | [102] |
| | Danish | rs10830963 | $P < 0.05$ | [103] |
| | | rs1387153 | $P < 0.05$ | [103] |
| | Saudi Arabians | rs1387153 | $P < 0.05$ | [104] |
| | | rs10830963 | $P < 0.05$ | [104] |
| PPARG | Asians | rs1801282 | $P = 0.011$ | [94] |
| | Caucasians | rs1801282 | $P < 0.05$ | [99] |
| | | rs3856806 | $P < 0.05$ | [99] |
| | Caucasians/Danish | rs1801282 | $P < 0.05$ | [98] |
| CDKAL1 | Koreans | rs7754840 | $P = 2.49 \times 10^{-13}$ | [102] |
| | Caucasians | rs7756992 | $P < 0.05$ | [99] |
| | Caucasian/Danish | rs7756992 | $P = 0.00017$ | [98] |
| | Iranians | rs7754840 | $P < 0.001$ | [105] |
| IRS1 | Scandinavians | IRS1 Arg972 | $P < 0.05$ | [106] |
| | Americans | Arg972 | $P < 0.05$ | [107] |
| | Saudi Arabians | rs1801278 | $P = 0.01$ | [108] |
| | Austro-Hungarians | rs7578326 | $P < 0.05$ | [109] |
| HMGA2 | Africans | rs138066904 | $P = 2.516 \times 10^{-9}$ | [96] |
| | Africa Americans | rs343092 | $P < 0.05$ | [96] |
| | Europeans | rs2258238 | $P < 0.05$ | [96] |
| IGF2BP2 | Koreans | rs4402960 | $P < 0.001$ | [110] |
| | Caucasians/Danish | rs4402960 | $P < 0.001$ | [111] |
| | Chinese | rs4402960 | $P < 0.001$ | [112] |

**Table 2** Most prevalent gestational diabetes mellitus genes across countries, ethnicity, and race in the world *(Continued)*

| Gene | Countries/ethnicity/race | Variants | Level of significance | References |
|---|---|---|---|---|
| *GCKR* | Malaysians | rs780094 | $P < 0.05$ | [113] |
| | American Caucasians | rs780094 | $P < 0.05$ | [114] |
| | Brazilians | rs780094 | $P < 0.05$ | [115] |
| *FGF21* | Iranians | Over expression of mRNA | $P < 0.001$ | [116] |
| | Chinese | Over expression of mRNA | $P < 0.001$ | [117] |
| | Australians | Over expression of mRNA | $P < 0.001$ | [118] |

## GDM testing, efficacy, and cost-effectiveness

Early detection of GDM is important to prevent its short- and long-term effects, especially the maternal progression to T2DM and fetal programming to DM later in life. Certain features such as BM1 above 30 kg/m$^2$ as well as previous GDM, baby birth weight of 4.5 kg or above, and macrosomia suggest a need for a GDM test [146]. Pregnant women with a family history of DM and ethnic groups with a high prevalence of DM such as Asian, Black, African-Caribbean or Middle Eastern should also consider the test [146, 147]. As stated earlier, there is no genetic testing procedure yet for GDM, however, two tests, namely glucose challenge test (GCT) and oral glucose tolerance test (OGTT) are frequently conducted at 24-28 weeks of pregnancy to diagnose GDM. The two tests can be done in succession known as 2-step screening, or OGTT alone can be done called 1-step screening.

In the 2-step screening, the GCT (otherwise known as a glucose screening test) is done first and entails testing the blood glucose one hour after drinking a sweet substance without fasting. If the blood glucose is 140 mg/dL (7.8 mmol/L) or higher, then an OGTT is necessary [148]. The OGTT measures blood glucose after 8-hour fasting, after which a glucose substance (75 g) is taken and blood glucose re-measured after 1, 2 and 3 hours. High glucose levels at any two or more of the blood test times suggest GDM [148]. Though the pathophysiology of GDM are similar with T2DM, a GCT of 200 mg/dL or more could indicate T2DM [148].

Relatively recently, serum levels of C-reactive protein (CRP) as well as glycated hemoglobin (HbA1c) and random blood sugar (RBS) are used as screening tools for GDM in the first trimester. CRP concentrations of about 6 mg/L or higher in undiluted serum samples are considered positive for GDM [149]. According to the International Association of the Diabetes and Pregnancy Study Group (IADPSG), the cutoff level of HbA1c is 6.5 % and RBS is 11.1 mmol (200mg/dL) [150].

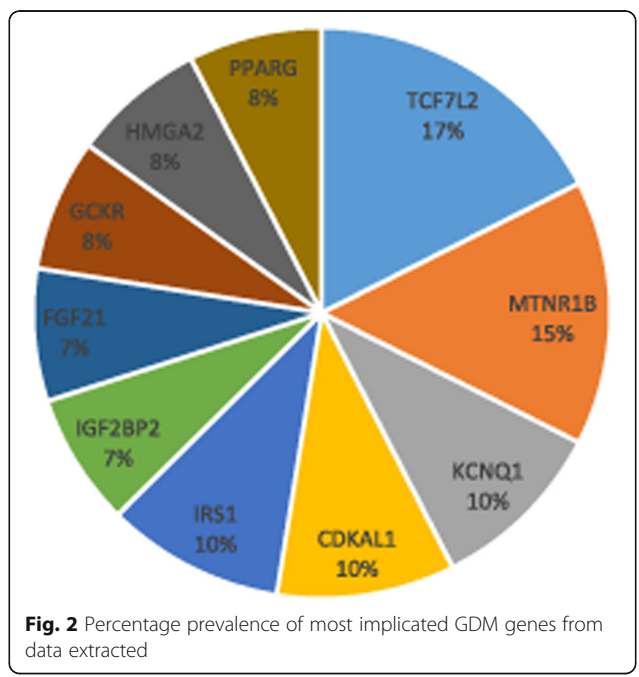

**Fig. 2** Percentage prevalence of most implicated GDM genes from data extracted

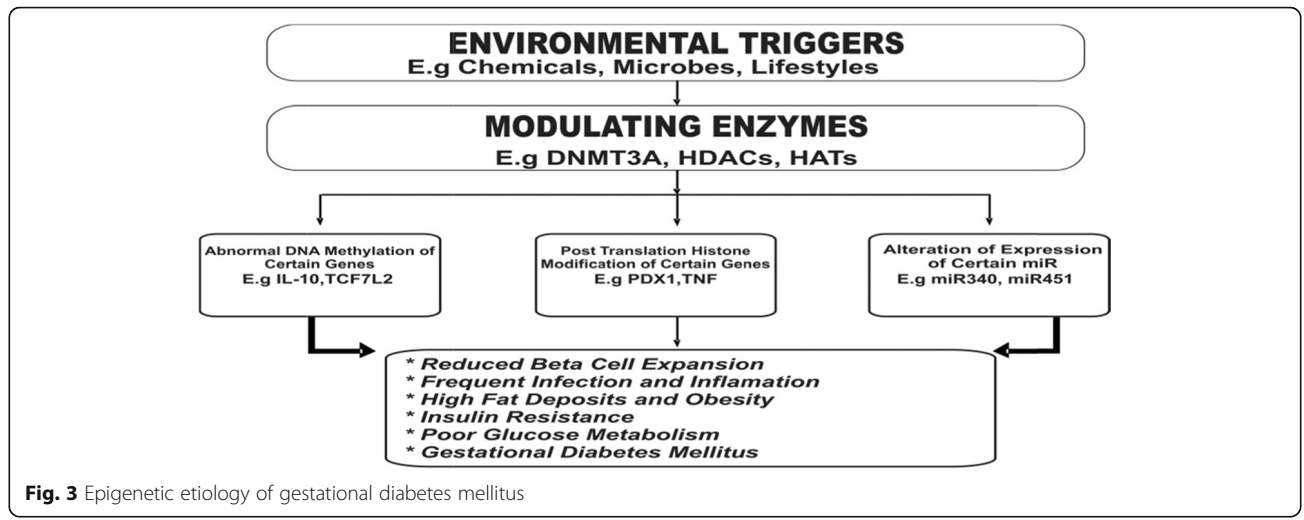

**Fig. 3** Epigenetic etiology of gestational diabetes mellitus

The cost-effectiveness of GDM testing is controversial because it depends on the region, race, screening tools, and methods. A systematic review by Fitria [151] reported that GDM testing and controlling is not effective in high-income countries. This could be due to the low prevalence of the disease in the region and the all-encompassing health care system. However, GDM testing could be worthwhile in countries with a high prevalence of GDM and nations with a poor healthcare system. For example, Marseille et al. [152] reported the cost-effectiveness of GDM testing using a devised model in Israel and India, which are known for high GDM incidence rates. The 2-step GTT is often recommended, however, a systematic review and meta-analysis by Saconne et al [153] showed no significant cost-effective difference between the two methods. The 2-step screening is also time-consuming and inconvenient, which may put off some patients [150, 154]. Glycated hemoglobin and RBS are simple GDM screening tools and are gaining acceptance worldwide, however, more awareness and understanding of the tools are necessary.

## Conclusion

Several articles reviewed showed that mutation and epigenetic modifications in certain genes can predispose humans to GDM. Most of the GDM candidate genes identified have also been implicated in the pathogenesis of T2DM, and both diseases share a common pathophysiology. The two metabolic disorders expressed oxidative stress-induced beta-cell dysfunction and insulin resistance through adiposity and obesity. One major difference between GDM and T2DM is that it resolves most times after delivery, however, it may progress to T2DM if not checked. This shows that GDM is a strong risk factor for T2DM, thus, its detection and management may reduce the prevalence of DM worldwide. Of the GDM candidate genes identified, the variants of *TCF7L2*, *MTNR1B*, *CDKAL1*, *IRS1* and *KCNQ1* are the most widespread, while some others are confined to certain ethnic groups. A genetic testing procedure can be developed around these genes to predict the likelihood of GDM or determine its genetic and epigenetic etiology. This will go a long way in stemming the incidence of DM worldwide.

### Abbreviations
CRP: C-reactive protein; DM: Diabetes mellitus; GCT: Glucose challenge test; GDM: Gestational diabetes mellitus; HbA1C: Glycated hemoglobin; IADPSG: International Association of the Diabetes and Pregnancy Study Group; MODY: Maturity onset diabetes of the young; NO₂: Nitrogen dioxide; OGTT: Oral glucose tolerance test; PM2.5: Particulate matter size 2.5; PM10: Particulate matter size 10; RBS: Random blood sugar; T1DM: Type 1 diabetes mellitus; T2DM: Type 2 diabetes mellitus

### Acknowledgements
None.

### Authors' contributions
TY conceptualized, did the literature search, drafted the manuscript, and did correspondence. TS designed the article and did the literature search. YA did the article selection and proofreading. AU did the article selection and proofreading. All authors have read and approved the manuscript.

### Funding
None.

### Availability of data and materials
Not applicable.

### Ethics approval and consent to participate
Not applicable.

### Consent for publication
Not applicable.

### Competing interests
The authors declare that they have no competing interests.

### Author details
[1]Federal University Birnin Kebbi, PMB 1157 Birnin Kebbi, Nigeria. [2]Department of Zoology and Environmental Biology, Olabisi Onabanjo University Ago-Iwoye, Ago-Iwoye, Ogun State, Nigeria. [3]Department of Biochemistry and Molecular Biology, Federal University Birnin Kebbi, Birnin Kebbi, Nigeria.

## Publisher's Note